\documentclass[%
 reprint,
 amsmath,amssymb,
 aps,
]{revtex4-2}

\usepackage[T1]{fontenc}
\usepackage{graphicx}
\usepackage{subfig}
\usepackage{caption}
\usepackage{hyperref}
\usepackage{amssymb,amsmath}
\usepackage{color}
\usepackage{units}
\usepackage{comment}
\usepackage{multirow}
\usepackage{makecell}
\usepackage{mathrsfs}
\usepackage[table,xcdraw]{xcolor}

\definecolor{grey}{gray}{.35} 
\definecolor{green}{rgb}{0.1, 0.5, 0.1}

\begin{document}
\title{Modularity, Hierarchical Flows and Symmetry of the Drosophila Connectome}

\author{Peter Grindrod \small{CBE}}
\email{grindrod@maths.ox.ac.uk}
\affiliation{Mathematical Institute, University of Oxford, Oxford, UK}

\author{Renaud Lambiotte}
\email{renaud.lambiotte@maths.ox.ac.uk}
\affiliation{Mathematical Institute, University of Oxford, Oxford, UK}

\author{Rohit Sahasrabuddhe}
\email{rohit.sahasrabuddhe@maths.ox.ac.uk}
\affiliation{Mathematical Institute, University of Oxford, Oxford, UK}

\date{\today}

\begin{abstract}
This report investigates the modular organisation of the Central region in the Drosophila connectome. We identify groups of neurones amongst which information circulates rapidly before spreading to the rest of the network using Infomap. We find that information flows along pathways linking distant neurones, forming modules that span across the brain. Remarkably, these modules, derived solely from neuronal connectivity patterns, exhibit a striking left-right symmetry in their spatial distribution as well as in their connections. We also identify a hierarchical structure at the coarse-grained scale of these modules, demonstrating the directional nature of information flow in the system.

\end{abstract}

\maketitle

\section{Introduction} 

In recent years, experimental advancements have dramatically enhanced our ability to produce detailed, large-scale maps of neural systems \cite{kulkarni2024towards}. Breakthroughs in electron microscopy and volumetric reconstructions allow researchers to explore increasingly complex organisms at the cellular level, mapping synaptic connections between neurones with precision \cite{dorkenwald2024neuronal, microns2021functional, shapson2024petavoxel}. 
With the bigger size and complexity of connectomic data, it is essential to develop methods to extract information from the myriad of connections, with the potential to guide future neuromorphic information processing concepts \cite{GrinBren}.
Network science offers such a powerful mathematical framework to connect structure and dynamics across scales and uncover fundamental principles underlying the architecture of neural systems \cite{newman2003structure, fornito2016fundamentals}.

In this short report, we extend a recent network analysis of the connectome of Drosophila \cite{lin2024network} by exploring its modular structure, which is known to be important in both structural and functional networks across organisms \cite{meunier2010modular}. 
Modular architecture highlights segregation and functional specialisation, and is associated with greater robustness, adaptivity, and evolvability of function. A modular representation \cite{fortunato2010community,lambiotte2021modularity} also creates a coarse-graining that reduces noise at the edge level while retaining structure at important scales.
We find that the Drosophila connectome is organised into hierarchical modules with information flowing asymmetrically between them, and whose spatial distribution and connectivity exhibit left-right symmetry.

\section{Results}

\subsection{Data description}

For our analysis, we use the \textit{Classification}, \textit{Connections}, and \textit{Coordinates} data made available by the FlyWire collaboration \cite{dorkenwald2024neuronal, schlegel2024whole}(\url{https://codex.flywire.ai}, accessed 11 November, 2024). We obtain the neurones of the Central super-class from \textit{Classification} and directed links between them from \textit{Connections}, where links with fewer than 5 synapses have been filtered out (though the methods here can be applied to a denser network with a lower threshold). \textit{Coordinates} provides the spatial coordinates for each neurone, and we take means where there are multiple observations for a neurone. After removing (very few) isolated neurones, the resulting network has 32,272 nodes with 849,981 directed unweighted edges. This network is a subgraph of the full connectome whose network properties including degree distributions, motifs, and rich-club organisation have been examined in   \cite{lin2024network}. 
We focus on the Central region of the brain, since that is where the main information processing may be taking place, following the preprocessing of sensory inputs. 
We are interested in the processing architecture which must have evolved for cognitive organisms to be compact and efficient, and to allow for fast responses to  early signals \cite{nicosia2013phase}.  

\subsection{Modularity}

\begin{figure*}[ht]
\centering
\includegraphics{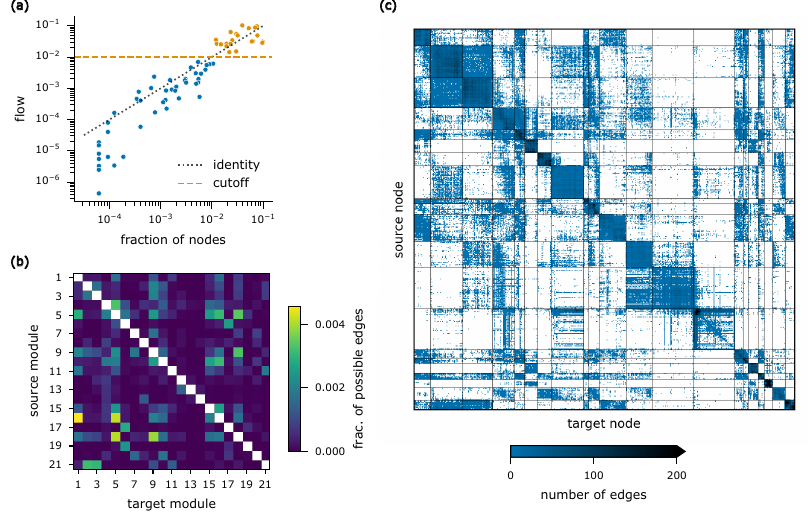}
\caption{\textbf{Modular structure.} \textbf{(a)} Flow  of each module versus the  fraction of nodes  within it. We focus our study on the 21 modules (orange markers) each with more than 1\% of the flow. \textbf{(b)} Fraction of possible edges observed between each pair of chosen modules. \textbf{(c)} Adjacency matrix of the subnetwork of 21 modules with nodes ordered first by module and then by x-coordinate. The black lines separate the modules. Since this is a large matrix, we capture the density of edges as a heatmap.}
\label{fig:modular_structure}
\end{figure*}

We identify groups of neurones amongst which information circulates rapidly by abstracting information flow as a random walk on the network. We use Infomap \cite{mapequation2024software} to find the network partition that minimises the description length of a random walk using a coding scheme that leverages modular structure \cite{rosvall2009map}. This flow-based approach focuses on the interplay of structure and dynamics, and can reveal markedly different modules than other methods \cite{rosvall2019different}, particularly in directed networks. We quantify the importance of a module by the \textit{flow} it captures, defined as the steady-state probability of a random walker being within it. We find 67 modules across orders of magnitude of flow and size, from which we pick the 21 which have more than 1\% of the flow for further study (Fig.\ref{fig:modular_structure}(a)). The sub-network of these modules (Fig.\ref{fig:modular_structure}(c)) has 28,887 (90\%) nodes, 761,005  (90\%) edges, and contains $0.94$ flow.


\subsection{Spatial distribution}
\begin{figure*}
    \centering
    \includegraphics{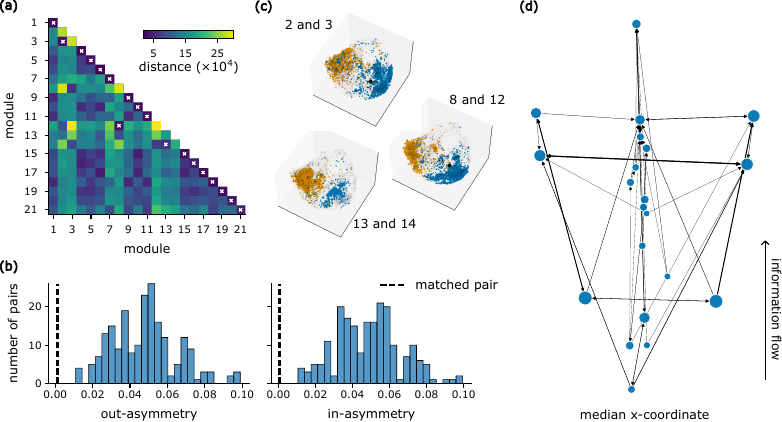}
    \caption{\textbf{Symmetry and hierarchy in modular structure.} \textbf{(a)} The (symmetric) matrix of Wasserstein 1-distance between each pair of modules where one is \textit{reflected}. White crosses mark best matches (see text). We identify 3 pairs of asymmetric best matches (off-diagonal). \textbf{(b)} Out and in-asymmetry (see text) of each pair of modules except for the matched pairs, which are marked by dashed black lines. \textbf{(c)} Spatial distribution of the asymmetric matched pairs. \textbf{(d)} Meso-scale network structure. Circles correspond to modules, with size indicating the number of nodes. We determine the location of a circle by the median $x$-coordinate of its neurones and by its position in the hierarchy created by information flow (see text). For visual clarity, we only plot edges in the backbone of the network, which we extract using a threshold of $0.9$ for the Disparity Filter \cite{serrano2009extracting}.}
    \label{fig:symmetry_hierarchy}
\end{figure*}

\begin{figure*}
    \centering
    \includegraphics{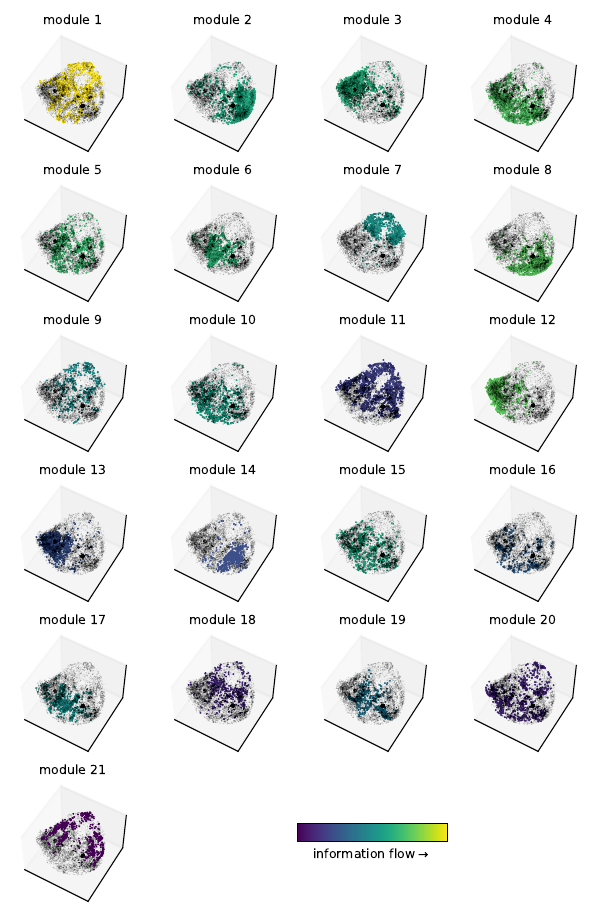}
    \caption{\textbf{Spatial locations of modules.} The colour of each module indicates its position in the hierarchy created by information flow (see text). We plot other neurones with light grey markers to give a sense of scale and location.}
    \label{fig:modules_in_space}
\end{figure*}

Many of the modules extend over large regions of the brain (Fig.\ref{fig:modules_in_space}), indicating that the flow of information connects distant neurones, creating groups that are distinct from spatial clusters (AMI$=0.35$ with the partition created by k-means clustering with 21 clusters). Treating the whole central region of the brain as a sampled point set in 3D, standard box counting methods \cite{BUCZKOWSKI199823}  estimate its Hausdorff dimension at around  2.5, whereas each of the 21 modules has a Hausdorff dimension estimate of  less than 2 (and a similar gap with nearest neighbour methods). Whilst only indicative, this  suggests modules within the partition are overlaid like layered strata within the three dimensional whole.

Strikingly, several modules appear visually to be symmetric across the left and right hemispheres. To study this further, we mean-centre the $x$ coordinate of every neurone and define the \textit{reflection} of a neurone at $(x,y,z)$ to be $(-x, y, z)$. Extending this notion to modules, we would expect symmetric modules to be similar to their reflections. We measure the spatial distance between modules (and their reflections) by considering them as probability distributions and using the Wasserstein 1-distance (also called Earth mover's distance) \cite{villani2021topics}. Intuitively, it measures the minimum cost of converting one distribution into the other, where the cost is amount the probability mass being moved times the distance. For computational ease, we discretise the space into cubes of 15000nm width, creating a $30 \times 20 \times 17$ histogram.

For each pair of modules $(\alpha,\beta)$, we measure the distance between $\alpha$ and the reflection of $\beta$ (identical to the distance between $\beta$ and the reflection of $\alpha$; Fig.\ref{fig:symmetry_hierarchy}(a)). The \textit{best match} of $\alpha$ is the module closest to its reflection. 15 modules are matched to themselves, indicating that they are symmetric. The 6 asymmetric modules are pairs of best matches -- 2 and 3, 8 and 12, and 13 and 14 (Fig.\ref{fig:symmetry_hierarchy}(c)). This remarkably clear matching up indicates that neurones are organised into spatially symmetric groups across the left and right hemispheres.


Each asymmetric module also appears visually to have symmetric connections to the rest of the network as their best match (Fig.\ref{fig:symmetry_hierarchy}(d)), suggesting that the pairs have similar structural roles in either hemisphere. To investigate this, we define the matrix $\mathbf{M}$ to contain the fraction of possible edges observed between modules (Fig.\ref{fig:modular_structure}(b)).
\begin{equation}
    M_{\alpha \beta} = 
    \begin{cases}
        \frac{n_{\alpha \beta}}{n_\alpha n_\beta}, &\text{if } \alpha \neq \beta\\
        \frac{n_{\alpha \alpha}}{n_\alpha(n_\alpha-1)}, &\text{if } \alpha = \beta,
    \end{cases}
\end{equation}
where $n_\alpha$ is the number of nodes in $\alpha$ and $w_{\alpha \beta}$ is the number of edges from $\alpha$ to $\beta$. Consider the matched pair of modules $(2,3)$. If they have similar roles, we would expect $M_{2\beta} \approx M_{3\beta}$, $M_{\beta 2} \approx M_{\beta 3}$ for symmetric modules $\beta$ and $M_{2\beta} \approx M_{3\beta^*}$, $M_{\beta 2} \approx M_{\beta^*3}$ for asymmetric matches $(\beta, \beta^*)$. To aid this comparison, we define $\mathbf{M'}$ as $\mathbf{M}$ with the rows and columns corresponding to the matched pairs swapped. We measure the out-asymmetry (resp. in-asymmetry) of a pair of modules $(\alpha, \beta)$, $\beta \neq \alpha$, as the L1 distance between the rows (resp. columns) of $\alpha$ in $\mathbf{M}$ and $\beta$ in $\mathbf{M'}$ (Fig.\ref{fig:symmetry_hierarchy}(b)). The values for matched pairs are nearly $0$ -- much lower than all other pairs, revealing that at a coarse-grained level, both spatial distribution and connectivity are symmetric across the hemispheres.

\subsection{Hierarchy}
Several neurones in the Drosophila connectome have a large difference between the number of inputs and outputs, and may serve as integrators and broadcasters \cite{lin2024network}. This can be thought of as a hierarchy created by the directionality of information flow. We investigate whether this hierarchy exists at a group level using SpringRank \cite{de2018physical}, an approach to assign real-valued ranks to nodes in directed networks, which is equivalent to the so-called Hodge decomposition and to trophic coherence \cite{mackay2020directed}. Intuitively, it infers a hierarchy where edges are likely to be directed `upwards' and connect nodes with similar ranks. We reveal a hierarchy in the network of modules (Fig.\ref{fig:symmetry_hierarchy}(d)), where information flows from `broadcaster' modules such as $21$ to `integrator' modules such as $1$. Using a null model that randomises the direction of edges while maintaining the total number between each pair of modules \cite{de2018physical}, we find that this hierarchy is statistically significant. We note that left-right symmetry is also apparent here, with the matched pairs having nearly identical ranks.

\section{Discussion}
In this report, we explore the modular structure of the Central region of the Drosophila connectome from a network science perspective. We find that information flows along paths connecting distant neurones, creating modules that extend across the brain. Despite being defined solely using the connections between neurones, the modules reveal left-right symmetry in the connectome, both in terms of the spatial distribution of the modules and their structural roles. We also find that the directionality of information flow creates hierarchical structure at the coarse-grained scale of modules.

The clear biological interpretation of an edge as a physical connection between neurones allows us to use network science methods  to obtain biologically interpretable results, and we believe that this work creates several avenues for future research. For instance, comparing the flow-based modules with groups created by spatial proximity or biological characteristics could create insights into how the brain processes information. Visual inspection of the adjacency matrix of the network suggests that modular structure might exist at different resolutions, making it important to investigate the structure of the connectome across scales. Further, while we primarily study inter-module connections, comparing the internal organisation of modules is crucial to understanding their biological function. Network science also offers ways to characterise structure other than groups such as roles, node rankings, motifs, and core-periphery organisation across scales. Finally, it would be insightful to explore the geometrical properties of the brain regions associated to each community, including their possible fractal nature.


\subsection*{Acknowledgments}
Authors are listed in alphabetical order of their last names.

We thank the FlyWire consortium \cite{schlegel2024whole,dorkenwald2024neuronal} for publicly releasing and maintaining the Drosophila connectome. We thank Sharon K. Whiteside for help with wrangling the data and Clive E. Bowman for advice.  We acknowledge the use of the open source libraries POT \cite{flamary2021pot}, Infomap \cite{mapequation2024software}, NetworkX \cite{SciPyProceedings_11}, and SpringRank \cite{de2018physical}. 

PG acknowledges support from EPSRC grant EP/Y007484/1. 
RL acknowledges support from the EPSRC grants EP/V013068/1, EP/V03474X/1 and EP/Y028872/1. RS acknowledges funding from the Mathematical Institute, University of Oxford.

The authors declare no conflicts of interest. For the purpose of open access, the authors have applied a CC BY public copyright licence to any Author Accepted Manuscript version arising from this submission.

\bibliographystyle{apsrev4-1}
\bibliography{references}

\end{document}